\documentclass[journal]{IEEEtran}
\usepackage{cite}
\usepackage{amsmath,amssymb,amsfonts}
\usepackage{algorithmic}
\usepackage{graphicx}
\usepackage{textcomp}
\usepackage{times}
\usepackage{bm}
\usepackage{multirow}
\usepackage{enumerate}
\usepackage{graphicx}
\usepackage{subfigure} 
\usepackage{multirow}
\usepackage{algorithmic}
\usepackage{algorithm}
\usepackage{caption}
\usepackage{color}
\usepackage{array}
\usepackage{epstopdf}
\definecolor{tabcolor}{rgb}{1,0,0}
\DeclareCaptionLabelSeparator{twospace}{.\ ~}
\newtheorem{theorem}{Theorem}

\def\BibTeX{{\rm B\kern-.05em{\sc i\kern-.025em b}\kern-.08em
		T\kern-.1667em\lower.7ex\hbox{E}\kern-.125emX}}
\begin{document} 
\title{Fluid Antenna for Mobile Edge Computing}
\author{Yiping Zuo, Jiajia Guo, \IEEEmembership{\normalsize {Member,~IEEE}}, Biyun Sheng, Chen Dai, \\ Fu Xiao, \IEEEmembership{\normalsize {Member,~IEEE}}, and Shi Jin, \IEEEmembership{\normalsize {Fellow,~IEEE}} 
\thanks{Yiping Zuo, Biyun Sheng, Chen Dai, and Fu Xiao are with the School of Computer Science, Nanjing University of Posts and Telecommunications, Nanjing 210023, China (Email: zuoyiping@njupt.edu.cn, biyunsheng@njupt.edu.cn, daichen@njupt.edu.cn, xiaof@njupt.edu.cn).} 
\thanks{Jiajia Guo and Shi Jin are with the National Mobile Communications Research Laboratory, Southeast University, Nanjing, 210096, China (email: jiajiaguo@seu.edu.cn, jinshi@seu.edu.cn).}
}
\maketitle
\begin{abstract}
In the evolving environment of mobile edge computing (MEC), optimizing system performance to meet the growing demand for low-latency computing services is a top priority. Integrating fluidic antenna (FA) technology into MEC networks provides a new approach to address this challenge. This letter proposes an FA-enabled MEC scheme that aims to minimize the total system delay by leveraging the mobility of FA to enhance channel conditions and improve computational offloading efficiency. By establishing an optimization problem focusing on the joint optimization of computation offloading and antenna positioning, we introduce an alternating iterative algorithm based on the interior point method and particle swarm optimization (IPPSO). Numerical results demonstrate the advantages of our proposed scheme compared to traditional fixed antenna positions, showing significant improvements in transmission rates and reductions in delays. The proposed IPPSO algorithm exhibits robust convergence properties, further validating the effectiveness of our method.  
\end{abstract}

\begin{IEEEkeywords}
Fluid antenna, mobile edge computing, antenna positioning, computation offloading, particle swarm 
\end{IEEEkeywords} 

\IEEEpeerreviewmaketitle
  
\section{Introduction} \label{sec:Introduction}
With the rapid development of wireless communications, mobile edge computing (MEC) has received widespread attention due to its potential in meeting low-latency and high-bandwidth requirements \cite{mao2017survey}. MEC technology brings data processing closer to the user end, thereby reducing the distance and time of data transmission in the network, and improving processing speed and efficiency \cite{zuo2023survey,abbas2017mobile}. However, with the explosive growth of the number of devices and data volume, existing MEC solutions face challenges in signal coverage and network capacity. 

Recently, fluid antenna (FA) \cite{wong2023fluid27}, also known as movable antenna \cite{zhu2023movable1}, as an emerging technology in the field of wireless communications, has attracted widespread attention for its ability to boost system performance through dynamic antenna adjustments. Studies explored the basic principles of FA technology, such as the study of a new spatial block correlation model for FA systems \cite{ramirez2024new}. Moreover, existing works focused on FA's performance in specific wireless communication scenarios, such as the FA-assisted multiple input multiple output (MIMO) communication systems \cite{ye2023fluid,zhu2024index} and the multi-user uplink communication systems based on FA \cite{hu2024fluid}. These studies demonstrate the potential of FA in improving spectral efficiency, reducing transmit power, and optimizing signal receiving quality. Meanwhile, the combination of FA with other emerging technologies, such as reconfigurable intelligent surfaces \cite{ghadi2024performance} and massive MIMO \cite{wong2022extra}, opens a new dimension in wireless communication system design.  

Given FA's inherent advantages, FA has the potential to address the challenges faced by MEC, such as reducing system delays and enhancing resource utilization efficiency. In this letter, we propose a novel FA-enabled MEC scheme, which aims to minimize the total system delay and improve MEC service quality by dynamically optimizing antenna positions and computing resource allocation. Specifically, this letter introduces a novel FA-enabled MEC scheme. Then, we formulate an optimization problem aimed at minimizing the total delay and design an alternating iterative algorithm based on the interior point method and particle swarm optimization (IPPSO) to find the optimal solution. Numerical experiments demonstrate that the proposed IPPSO-based algorithm has good convergence. Comparing with two baseline schemes, the proposed FA-enabled MEC scheme has significant advantages in reducing the total system delay.

\section{System Model} \label{sec:SystemModel}
\begin{figure}[!t]
	\centering    
	\includegraphics[width=1.0\linewidth]{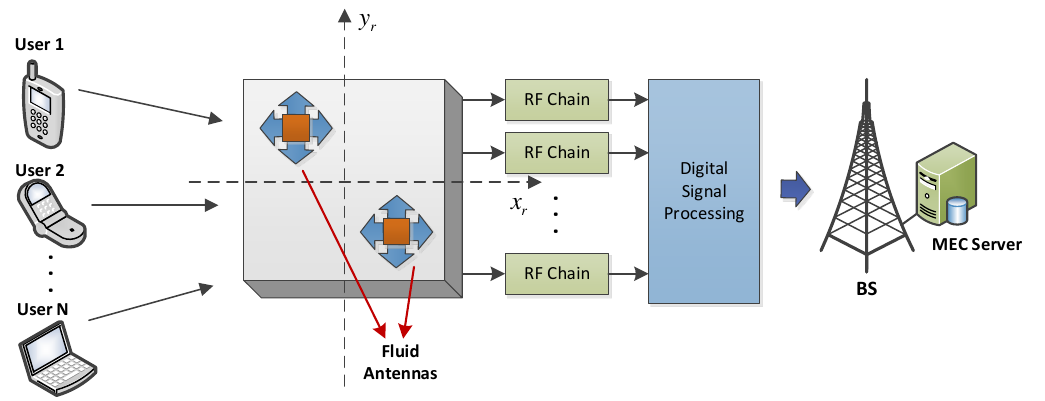}   
	\caption{The FA-enabled MEC networks}
	\label{fig:SystemModel}
\end{figure}

Fig.~\ref{fig:SystemModel} illustrates an FA-enabled MEC network. This network comprises $N$ single-antenna users and a FA-enabled BS. The BS has a MEC server managed by the cloud service provider of the core network. Notably, antennas on users remain stationary, whereas $M$ FAs on the BS are mobile within a local domain. The set of all users is denoted by ${\cal N} = \left\{ {1,2, \cdots ,N} \right\}$. Let ${\cal M} = \left\{ {1,2, \cdots ,M} \right\}$ denote the set of all FAs at the BS. This local domain can be viewed as a rectangle in a two-dimensional coordinate system, denoted as ${{\cal D}_r}$. Each FA is connected to the radio frequency (RF) chain via a flexible cable, thereby enhancing the channel conditions between the BS and users. We consider space-division multiple access for users concurrently communicating with the BS in the uplink transmission. Hence, we have an assumption that the number of users does not surpass the number of FAs at the BS, i.e., $N \le M$. The position of the $m$-th receive FA at the BS is defined as ${{\bf{d}}_m} = {\left[ {{x_m},{y_m}} \right]^{\rm{T}}} \in {{\cal D}_r}$ for $m \in {\cal M}$.

\subsection{Communication Model}\label{subsec:CommunicationModel}
We consider the uplink transmission from users to the BS. Then, the received signal ${\mathbf{y}} \in {\mathbb{C}^{N \times 1}}$ at the BS can be expressed as
\begin{equation}\label{ReceivedSignal_y} 
{\mathbf{y}} = {{\mathbf{W}}^H}{\mathbf{H}}\left( {\mathbf{d}} \right){{\mathbf{P}}^{{1 \mathord{\left/
				{\vphantom {1 2}} \right.
				\kern-\nulldelimiterspace} 2}}}{\mathbf{s}} + {{\mathbf{W}}^H}{\mathbf{n}},
\end{equation}
where ${\mathbf{W}} = \left[ {{{\mathbf{w}}_1},{{\mathbf{w}}_2}, \cdots ,{{\mathbf{w}}_N}} \right] \in {\mathbb{C}^{M \times N}}$ represents the receive combining matrix at the BS with ${{\mathbf{w}}_n}$ being the combining vector for the transmitted signal of user $n$. ${\mathbf{H}}\left( {\mathbf{d}} \right) = \left[ {{{\mathbf{h}}_1}\left( {\mathbf{d}} \right),{{\mathbf{h}}_2}\left( {\mathbf{d}} \right), \cdots ,{{\mathbf{h}}_N}\left( {\mathbf{d}} \right)} \right] \in {\mathbb{C}^{M \times N}}$ is the multiple-access channel matrix from all $N$ users to the $M$ FAs at the BS with ${\bf{d}} = {\left[ {{\bf{d}}_1^T,{\bf{d}}_2^T, \cdots ,{\bf{d}}_M^T} \right]^T}$ denoting the antenna position vector for FAs. ${{\bf{P}}^{{1 \mathord{\left/ {\vphantom {1 2}} \right. \kern-\nulldelimiterspace} 2}}} = {\rm{diag}}\left\{ {\sqrt {{p_1}} ,\sqrt {{p_2}} , \cdots ,\sqrt {{p_N}} } \right\} \in {\mathbb{C}^{N \times N}}$ denotes the power scaling matrix, where ${p_n}$ is the transmit power of user $n$. ${\mathbf{s}} = {[{s_1},{s_2}, \cdots ,{s_N}]^T} \in {\mathbb{C}^{N \times 1}}$ is the transmit signal vector of all users, and $s_n$ denotes the transmitted signal of user $n$ with normalized power, i.e., $\mathbb{E}\left( {{\mathbf{s}}{{\mathbf{s}}^H}} \right) = {{\mathbf{I}}_N}$. Additionally, ${\mathbf{n}} = {\left[ {{n_1},{n_2}, \cdots ,{n_M}} \right]^T} \sim \mathcal{C}\mathcal{N}\left( {{\mathbf{0}},{\sigma ^2}{{\mathbf{I}}_M}} \right)$ denotes the zero-mean additive white Gaussian noise with ${{\sigma ^2}}$ denoting the average noise power, in which $n_m$ is the noise at the $m$-th FA antenna at the BS.

The channel vector between each user and the BS is determined by the propagation environment and the location of FAs. Since the moving area of FAs at the BS is much smaller than the signal propagation distance, we assume that the condition of far field is satisfied between users and the BS. For each user, the angles of arrival (AoAs) and the amplitudes of the complex path coefficients for multiple channel paths remain constant for different positions of FAs. This implies that only a phase change occurs for multiple channels in the receiving area. Let $L_n$ denote the total number of receive channel paths at the BS from user $n$. The set of all channel paths at the BS is denoted by ${\mathcal{L}_n} = \left\{ {1,2, \cdots ,{L_n}} \right\}$. We adopt a channel model based on the field response, then the channel vector between user $n$ and the BS is given as follow
\begin{equation}\label{ChannelVector_hn} 
{{\mathbf{h}}_n}\left( {\mathbf{d}} \right) = {\mathbf{F}}_n^H\left( {\mathbf{d}} \right){{\mathbf{G}}_n},
\end{equation}
where ${{\mathbf{F}}_n}\left( {\mathbf{d}} \right) = \left[ {{{\mathbf{f}}_n}\left( {{{\mathbf{d}}_1}} \right),{{\mathbf{f}}_n}\left( {{{\mathbf{d}}_2}} \right), \cdots ,{{\mathbf{f}}_n}\left( {{{\mathbf{d}}_M}} \right)} \right] \in {\mathbb{C}^{{L_n} \times M}}$ represents the field-response matrix at the BS with ${{\mathbf{f}}_n}\left( {{{\mathbf{d}}_m}} \right) = {\left[ {{e^{j\frac{{2\pi }}{\lambda }{\rho _{n,1}}\left( {{{\mathbf{d}}_m}} \right)}},{e^{j\frac{{2\pi }}{\lambda }{\rho _{n,2}}\left( {{{\mathbf{d}}_m}} \right)}}, \cdots ,{e^{j\frac{{2\pi }}{\lambda }{\rho _{n,{L_n}}}\left( {{{\mathbf{d}}_m}} \right)}}} \right]^T}$ denoting the field-response vector of the received channel paths between user $n$ and the $m$-th FA at the BS. In ${{\mathbf{f}}_n}\left( {{{\mathbf{d}}_m}} \right)$, ${\rho _{n,l}}\left( {{{\mathbf{d}}_m}} \right) = {x_m}\sin {\theta _{n,l}}\cos {\phi _{n,l}} + {y_m}\cos {\theta _{n,l}}$ is the phase difference in the signal propagation for the $l$ path from user $n$ between the position of the $m$-th FA and the reference point denoted by ${{\mathbf{d}}_0} = {\left[ {0,0} \right]^T}$, where ${\theta _{n,l}}$ and ${\phi _{n,l}}$ represent the elevation and azimuth AoAs for the $l$-th receive path between user $n$ and the BS. The path-response vector is denoted as ${{\mathbf{G}}_n} = {\left[ {{g_{n,1}},{g_{n,2}}, \cdots ,{g_{n,{L_n}}}} \right]^T}$, representing the coefficients of multi-path responses from user $n$ to the reference point in the receive region.

\subsection{Computation Offloading Model}\label{subsec:ComputationOffloadingModel}
All user run the federated learning (FL) training tasks, some tasks are offloaded to the MEC server at the BS due to limited computing resources of users. We let ${{\beta _n}}$ denote the proportion of FL training datasets offloaded to the MEC server from user $n$. Furthermore, both at the user end and the MEC server, we employ the stochastic gradient descent (SGD) optimization algorithm to train the FL model. 

In the local model training scheme $\left( {{\beta _n} = 0} \right)$, user $n$ only trains FL tasks locally. Therefore, the local training latency for user $n$ is given by $T_n^{loc} = \frac{{{C_n}{D_n}{\varepsilon _n}{\iota _n}}}{{f_n^{loc}}}$, where ${C_n}$ is regarded as the number of CPU cycles required to precess a single data sample for user $n$, ${\iota _n}$ represents the number of iterations of the SGD algorithm on user $n$, and $f_n^{loc}$ is the local CPU working frequency of user $n$. $B_n = D_n\varepsilon _n$ is the mini-batch sizes of user $n$, where $\varepsilon _n^{\left( k \right)} \in \left( {0,1} \right]$ indicates the mini-batch size ratio that refers to the ratio of the mini-batch size used in the local training to the total dataset size. Once the FL task is trained, mobile users will upload their local model parameters to the nearby MEC server. So, we obtain the latency taken for user $n$ to upload the local model parameters to the MEC server as $T_n^{up} = \frac{{{V_n}}}{{{R_n}}}$, where $V_n$ is the size of the local model parameters of user $n$. For simplicity, we assume that the size of model parameters is usually a constant multiple of the data set size denoted as $V_n = vD_n$, where $v$ is a constant. The transmission rate from user $n$ to the MEC server is expressed as
\begin{equation}\label{equ:TransmissionRate_Rn} 
{R_n} = {\log _2}\left( 1 + {\frac{{{{\left| {{\mathbf{w}}_n^H{{\mathbf{h}}_n}\left( {\mathbf{d}} \right)} \right|}^2}{p_n}}}{{\sum\nolimits_{k \in \mathcal{N},k \ne n} {{{\left| {{\mathbf{w}}_k^H{{\mathbf{h}}_k}\left( {\mathbf{d}} \right)} \right|}^2}{p_k} + \left\| {{{\mathbf{w}}_n}} \right\|_2^2{\sigma ^2}} }}} \right).
\end{equation}
We assume that the BS employs the widely used linear zero-forcing (ZF) detector for processing multiple signals, owing to its low implementation complexity \cite{hu2023movable}. Based on this assumption, the receive combining matrix ${{\mathbf{w}}_n}$ is accordingly expressed as
\begin{equation}\label{equ:wn_zf} 
{{\mathbf{w}}_n} = {{\mathbf{h}}_n}\left( {\mathbf{d}} \right){\left( {{{\mathbf{h}}_n}{{\left( {\mathbf{d}} \right)}^H}{{\mathbf{h}}_n}\left( {\mathbf{d}} \right)} \right)^{ - 1}}.
\end{equation}
Substituting the equations (\ref{equ:wn_zf}) into (\ref{equ:TransmissionRate_Rn}), we can obtain
\begin{equation}\label{equ:TransmissionRate_Rn_v2} 
{R_n} = {\log _2}\left( 1 + {\frac{{{p_n}}}{{\left\| {{{\mathbf{w}}_n}} \right\|_2^2{\sigma ^2}}}} \right).
\end{equation}

In the fully offloaded model training scheme $\left( {{\beta _n} = 1} \right)$, user $n$ offloads all datasets to the MEC server, and then trains the FL model on the MEC server. As a result, the transmission latency required to transfer the datasets from user $n$ to the MEC server is calculated as $T_n^{off} = \frac{{{D_n}}}{{{R_n}}}$. Then, the execution time of model training for user $n$ on the MEC server is expressed as $T_n^{exe} = \frac{{{C_M}{D_n}{\varepsilon _M}{\iota _M}}}{{f_n^M}}$,
where ${C_M}$ is regarded as the number of CPU cycles required to precess a single data sample for the MEC server, ${\iota _M}$ represents the number of iterations of the SGD algorithm on the MEC server, and $\varepsilon _M \in \left( {0,1} \right]$ indicates the mini-batch size ratio on the MEC server. $f_n^{M}$ refers to the CPU frequency of user $n$ from the MEC server.

Combining the above two model training schemes, under the mixed model training scheme, user $n$ locally trains a certain proportion of FL tasks, and offloads the remaining tasks to the corresponding the MEC server. Therefore, the total training latency for user $n$ is defined as the maximum one of the two model training parts, which is calculated as 
\begin{equation}\label{equ:Tn_v2} \footnotesize 
\begin{gathered}
{T_n} = \left( {1 - {\beta _n}} \right)\left( {\frac{{{C_n}{D_n}{\varepsilon _n}{\iota _n}}}{{f_n^{loc}}} + \frac{{{V_n}}}{{{R_n}}}} \right) + {\beta _n}\left( {\frac{{{D_n}}}{{{R_n}}} + \frac{{{C_M}{D_n}{\varepsilon _M}{\iota _M}}}{{f_n^M}}} \right).  
\end{gathered}
\end{equation}
Then, the sum latency of all users can be expressed as $T = \sum\nolimits_{n \in \mathcal{N}} {{T_n}}$.

\section{Problem Formulation and Analysis} \label{sec:ProblemFormulation}
This letter aims to minimize the sum latency of the proposed FA-enabled MEC networks by jointly optimizing the offloading ratio $\bm{\beta} = \left\{ {{\beta _1}, \cdots ,{\beta _N}} \right\}$, CPU frequency ${{\mathbf{f}}^M} = \left\{ {f_1^M, \cdots ,f_N^M} \right\}$, and antenna positioning ${\bf{d}} = {\left[ {{\bf{d}}_1^T, \cdots ,{\bf{d}}_M^T} \right]^T}$. As a result, the minimum latency optimization problem is formulated as
\begin{equation}\label{equ:P1} 
\begin{array}{l}
{\cal P}1:\mathop {\min }\limits_{{\bf{d}},\bm{\beta},{{\bf{f}}^M}} \;T\\
\;\;\;\;\;\;{\rm{s}}{\rm{.t}}{\rm{.}}\;{\rm{C1:}}{{\bf{d}}_m} \in {{\cal D}_r},\forall m \in {\cal M},\\
\;\;\;\;\;\;\;\;\;\;\;\;{\rm{C2:}}{\beta _n} \in \left\{ {0,1} \right\},\forall n \in {\cal N},\\
\;\;\;\;\;\;\;\;\;\;\;\;{\rm{C3:}}\sum\nolimits_{n \in {\cal N}} {f_n^M}  \le {{\bar f}_M},\\
\;\;\;\;\;\;\;\;\;\;\;\;{\rm{C4:}}\left\| {{{\bf{d}}_m} - {{\bf{d}}_k}} \right\| \ge {d_0},m \ne k,\forall m \in {\cal M},\\
\;\;\;\;\;\;\;\;\;\;\;\;{\rm{C5:}} {T_n} \le {{\bar T}_n},\forall n \in {\cal N},
\end{array}
\end{equation}
where constrain ${\rm{C1}}$ restricts the positioning of FA within the given region ${{\cal D}_r}$. Constrain ${\rm{C2}}$ sets the offloading ratio from each user to the MEC server as the binary variable. Constraint ${\rm{C3}}$ guarantees the CPU frequency of all users from the MEC server to not exceed the maximum CPU frequency ${{\bar f}_M}$. Constraint ${\rm{C4}}$ ensures that the distance between FAs is not less than ${d_0}$. Constraint ${\rm{C5}}$ indicates that the latency of each user does not exceed a predetermined maximum latency ${{\bar T}_n}$. By observing the optimization problem ${\cal P}1$, we can conclude that the problem ${\cal P}1$ is a non-convex optimization problem, also referred as an NP-hard problem. Generally, the efficient resolution of such problems poses considerable difficulty. Moreover, the existence of a high-dimensional search space compounds the challenge of attaining the global optimal solution. To address this problem, we can implement the following transformation and relaxation:
\begin{itemize}
\item[1)] \textbf{\emph{Variable Relaxation}} \\
We first perform continuous relaxation of the target variable ${\beta _n}$, as follows
\begin{equation}\label{Relaxation_Betan} 
0 \le {\beta _n} \le 1.
\end{equation}
Find the optimal solution of this continuous variable. Then we can obtain the optimal binary offloading strategy $\beta _n^*$ from user $n$ to the MEC server by passing the continuous solution through the threshold judgment.
\end{itemize}

\begin{itemize}
	\item[2)] \textbf{\emph{Problem Decomposition}} \\
	Based on the above variable relaxation, the original optimization problem ${\cal P}1$ can be equivalent to ${\cal P}2$, as follows
	\begin{equation}\label{equ:P2} 
	\begin{array}{l}
	{\cal P}2:\mathop {\min }\limits_{{\bf{d}},\bm{\beta},{{\bf{f}}^M}} \;T\\
	\;\;\;\;\;\;{\rm{s}}{\rm{.t}}{\rm{.}}\;{\rm{C1}},{\rm{C3}}-{\rm{C5}},\\
	\;\;\;\;\;\;\;\;\;\;\;\;{\rm{C2':0}} \le {\beta _n} \le 1,\forall n \in {\cal N}.
	\end{array}
	\end{equation}	
	To simplify the solution process of problem ${\cal P}2$, we consider decoupling ${\cal P}2$ into the following sub-problems ${\cal P}2-1$ and ${\cal P}2-2$, as follows
	
	\begin{equation}\label{equ:P2-1} 
	{\cal P}2 - 1:\mathop {\min }\limits_{\bm{\beta},{{\bf{f}}^M}} \;T, \text{s.t. } {\rm{C2'}}, {\rm{C3}}, {\rm{C5}},
	\end{equation}
	\begin{equation}\label{equ:P2-2} 
	\mathcal{P}2 - 2:\mathop {\min }\limits_{\mathbf{d}} \;T, \text{s.t. } {\rm{C1}}, {\rm{C4}}, {\rm{C5}}.
	\end{equation}

\end{itemize}

Next, we first discuss the sub-problem ${\cal P}2 - 1$ and can obtain the optimal solution to this convex optimization problem, as shown in the following theorem.
\begin{theorem}\label{Theorem:p2-1}
The sub-problem ${\cal P}2 - 1$ is a convex optimization problem.
\end{theorem}

\begin{IEEEproof}\label{ProofTheorem:p2-1}
After the above target variable relaxation, constraint ${\rm{C2'}}$ is a closed set and constraint ${\rm{C3}}$ is a convex set. Obviously, the utility function $T$ is a continuous function in convex sets of constraints ${\rm{C2'}}$, ${\rm{C3}}$, and ${\rm{C5}}$. Then, we take the second-order derivatives of $T$ with respect to ${\beta _n}$ and $f_n^M$ to obtain the Hessian matrix, which can be written as follows	
\begin{equation}\label{equ:HessianMatrix} 
{\mathcal{H}_n} = \left[ {\begin{array}{*{20}{c}}
	{\frac{{\partial {T^2}}}{{\partial \beta _n^2}}}&{\frac{{\partial {T^2}}}{{\partial {\beta _n}f_n^M}}} \\ 
	{\frac{{\partial {T^2}}}{{\partial f_n^M{\beta _n}}}}&{\frac{{\partial {T^2}}}{{\partial {{\left( {f_n^M} \right)}^2}}}} 
	\end{array}} \right] = \left[ {\begin{array}{*{20}{c}}
	0&0 \\ 
	0&{\frac{{2{C_M}{D_n}{\varepsilon _M}{\iota _M}{\beta _n}}}{{{{\left( {f_n^M} \right)}^3}}}} 
	\end{array}} \right]\underset{\raise0.3em\hbox{$\smash{\scriptscriptstyle}$}}{ \succeq } 0.
\end{equation}
Therefore, the objective function $T$ is a convex function with respect to ${\beta _n}$ and $f_n^M$. Accordingly, and the sub-problem ${\cal P}2 - 1$ is a convex optimization problem. This completes the proof.
\end{IEEEproof}
Based on Theorem~\ref{Theorem:p2-1}, the sub-problem ${\cal P}2 - 1$ is a convex optimization problem. Traditional optimization methods, such as the interior point method and the standard gradient projection method, can be used to find the suboptimal offloading ratio and CPU frequency strategies of ${\cal P}2 - 1$.

Then, we discuss the sub-problem ${\cal P}2 - 2$. If the optimal solution of ${\cal P}2 - 2$ is directly searched, the solution solution ${{\cal D}_r}$ of antenna positioning is typically large, which can result in excessively high computational complexity. To address this challenge, particle swarm optimization (PSO) \cite{xiao2023multiuser} is introduced as an effective approach.

In the PSO-based algorithm, we set the number of particles as $I$ with positions ${\mathcal{D}^{\left( 0 \right)}} = \left\{ {{\mathbf{d}}_1^{\left( 0 \right)},{\mathbf{d}}_2^{\left( 0 \right)}, \cdots ,{\mathbf{d}}_I^{\left( 0 \right)}} \right\}$ and velocities ${\mathcal{V}^{\left( 0 \right)}} = \left\{ {{\mathbf{v}}_1^{\left( 0 \right)},{\mathbf{v}}_2^{\left( 0 \right)}, \cdots ,{\mathbf{v}}_I^{\left( 0 \right)}} \right\}$, where each particle indicates a possible solution for the antenna positioning at the BS. Here, ${\mathbf{d}}_i^{\left( 0 \right)}$ can be expressed as
\begin{equation}\label{equ:EachParticle_di0} 
{\mathbf{d}}_i^{\left( 0 \right)} = {\left[ {\underbrace {x_{i,1}^{\left( 0 \right)},y_{i,1}^{\left( 0 \right)}}_{{\text{FA}}\;{\text{1}}},\underbrace {x_{i,2}^{\left( 0 \right)},y_{i,2}^{\left( 0 \right)}}_{{\text{FA}}\;2}, \cdots ,\underbrace {x_{i,M}^{\left( 0 \right)},y_{i,M}^{\left( 0 \right)}}_{{\text{FA}}\;M}} \right]^T},
\end{equation}
where the initial position of each particle $i$ follows a uniform distribution, denoted as $x_{i,m}^{\left( 0 \right)},y_{i,m}^{\left( 0 \right)} \sim \mathcal{U}\left[ { - A,A} \right]$ for $1 \le i \le I$ and $1 \le m \le M$. We set the feasible area for antenna movement to be ${\mathcal{D}_r} \in \left[ { - A,A} \right] \times \left[ { - A,A} \right]$. To satisfy constraint ${\rm{C1}}$, we need to constrain the position of particle $i$ and antenna $m$ within the feasible area. The specific mathematical expression is as follows
\begin{equation}\label{equ:ximyim} 
\left\{ \begin{gathered}
{x_{i,m}} = \max \left( {\min \left( {{x_{i,m}},A} \right), - A} \right), \hfill \\
{y_{i,m}} = \max \left( {\min \left( {{y_{i,m}},A} \right), - A} \right). \hfill \\ 
\end{gathered}  \right.
\end{equation}

Each particle updates its speed and new position according to the following formula
\begin{equation}\label{equ:ViDi} 
\begin{gathered}
{{\mathbf{v}}_i}\left( {t + 1} \right) = w{{\mathbf{v}}_i}\left( {t + 1} \right) + {c_1}{r_1}\left( {{{\mathbf{d}}_{popt}} - {{\mathbf{d}}_i}\left( t \right)} \right) + \hfill\\
\;\;\;\;\;\;\;\;\;\;\;\;\;\;\;\;\;\;\;{c_2}{r_2}\left( {{{\mathbf{d}}_{gopt}} - {{\mathbf{d}}_i}\left( t \right)} \right), \hfill \\
{{\mathbf{d}}_i}\left( {t + 1} \right) = {{\mathbf{d}}_i}\left( t \right) + {{\mathbf{v}}_i}\left( {t + 1} \right), 1 \le i \le I, \hfill \\ 
\end{gathered} 
\end{equation}
where ${{\mathbf{d}}_{popt}}$ is the personal optimal solution and ${{\mathbf{d}}_{gopt}}$ represents the global optimal solution. $c_1$ and $c_2$ are positive learning factors, and $r_1$ and $r_2$ are uniformly distributed random numbers between 0 and 1. The inertia weight, $w$, is employed using the linear decrease method, which is expressed as
\begin{equation}\label{equ:w_linerdecrease} 
w = {w_{\max }} - \frac{{\left( {{w_{\max }} - {w_{\min }}} \right)t}}{\cal T},
\end{equation}
where ${w_{\max }}$ and ${w_{\min }}$ are the inertia weights of the maximum and minimum values respectively. $t$ is the current iteration number and $\cal T$ is the total number of the iteration planned for the PSO process.

To satisfy constraints ${\rm{C4}}$ and ${\rm{C5}}$, we define the penalty function as
\begin{equation}\label{equ:PenaltyFunction} \footnotesize
{\mathcal{P}_i}\left( {\mathbf{d}} \right) = {\tau _1}{\sum\limits_i^{{N_v}} {\left( {T_n^{\left( i \right)}\left( {\mathbf{d}} \right) - {{\bar T}_n}} \right)} ^2} + {\tau _2}{N_d}, 1 \le i \le I,
\end{equation}
where both ${\tau _1}$ and ${\tau _2}$ represent the penalty coefficient, which is used to adjust the severity of the penalty. $T_n^{\left( i \right)}$ represents the actual delay when constraint ${\rm{C5}}$ is violated. ${N_v}$ and ${N_d}$ represent the number of illegal constraints ${\rm{C4}}$ and ${\rm{C5}}$ respectively. Among them, ${N_d} = \sum\limits_{m = 1}^{M - 1} {\sum\limits_{k = m + 1}^M {\delta \left( {d\left( {{{\mathbf{d}}_m},{{\mathbf{d}}_k}} \right) < {d_0}} \right)} }$, where $\delta \left(  \cdot \right)$ is an indicator function, which is 1 when the condition in the brackets is true, otherwise it is 0. Based on this given penalty function, we define the fitness function of the PSO-based algorithm 
\begin{equation}\label{equ:FitnessFunction} 
{\mathcal{F}_i}\left( {\mathbf{d}} \right) = {T_i}\left( {\mathbf{d}} \right) + {\mathcal{P}_i}\left( {\mathbf{d}} \right), 1 \le i \le I.
\end{equation}

Building on the foregoing discussion and analysis, we develop an IPPSO-based alternating iteration algorithm aimed at addressing problem ${\cal P}2$. For a more detailed step-by-step process, the proposed IPPSO-based alternating iterative algorithm is summarized in Algorithm~\ref{Algorithm_IPPSO}. The complexity of the entire Algorithm~\ref{Algorithm_IPPSO} is $\mathcal{O}\left( {\mathcal{K} \cdot \left( {{{\left( {2N + 1} \right)}^3} + \mathcal{T}I\left( {{M^2} + M + N} \right)} \right)} \right)$, where ${\mathcal{K}}$ denotes the number of iterations of the outer loop, and ${\mathcal{T}}$ signifies the number of iterations of the inner loop.

\begin{algorithm}[!t] \scriptsize
	\caption{IPPSO-based Alternating Iterative Algorithm} \label{Algorithm_IPPSO}                                
	\setcounter{algorithm}{1}
	\begin{algorithmic}[1]
		\STATE{\textbf{Input:} The initial data set $(M, N, L, \lambda, D_n, {f_n^{loc}},{\theta _{n,l}}, {\phi _{n,l}}$, ${{\bar f}_M}, {{\bar T}_n}, p_n, g_{n,L_{n}}, {\varepsilon _n}, {\iota _n}, {\varepsilon _M}, {\iota _M}, v, d_{0})$ for $n \in {\cal N}$, $m \in {\cal M}$, and $l\in {{\cal L}_n}$.}
		\STATE {\textbf{Initialization:} Initialize target variables ${\bf{d}}$, $\bm{\beta}$, and ${{\bf{f}}^M}$, the $N$ particles with positions ${\mathcal{D}^{\left( 0 \right)}}$ and velocities ${\mathcal{V}^{\left( 0 \right)}}$, the individual optimal position $\mathbf{d}_{popt}$ and the global optimal position $\mathbf{d}_{gopt}$.}
		\FOR {Outer-loop iteration $k$=1 to $\cal K$}
		\STATE {Fix ${\bf{d}}$, obtain the optimal solutions $\bm{\beta}^{*}$ and ${{\bf{f}}^{M*}}$ of the sub-problem ${\cal P}2-1$ using the interior-point algorithm.}
		\STATE {Update $\bm{\beta}$ and ${{\bf{f}}^{M}}$.}
		\FOR {Innter-loop iteration $t$=1 to $\cal T$}
		\STATE {Calculate the inertia weight $w$ according to (\ref{equ:w_linerdecrease}).}
		\FOR {Each particle $i$ = to $I$}
		\STATE {Fix $\bm{\beta}$ and ${{\bf{f}}^{M}}$, calculate the fitness function value ${\mathcal{F}_i}\left( {\mathbf{d}} \right)$ according to (\ref{equ:FitnessFunction}).}
		\STATE {Update the velocity and position of each particle according to (\ref{equ:ViDi}).}
		\IF{${\mathcal{F}_i}\left( {{{\mathbf{d}}^{\left( t \right)}}} \right) < {\mathcal{F}_i}\left( {{\mathbf{d}}_{popt}^{\left( t \right)}} \right)$}
		\STATE {Update ${\mathbf{d}}_{popt}^{\left( t \right)} \leftarrow {{\mathbf{d}}^{\left( t \right)}}$.}
		\ENDIF
		\IF{${\mathcal{F}_i}\left( {{{\mathbf{d}}^{\left( t \right)}}} \right) < {\mathcal{F}_i}\left( {{\mathbf{d}}_{gopt}^{\left( t \right)}} \right)$}
		\STATE {Update ${\mathbf{d}}_{gopt}^{\left( t \right)} \leftarrow {{\mathbf{d}}^{\left( t \right)}}$.}
		\ENDIF
		\STATE {Perform boundary processing on antenna positions according to (\ref{equ:ximyim}).}
		\ENDFOR
		\ENDFOR
		\STATE {Update ${\mathbf{d}} \leftarrow {{\mathbf{d}}_{gopt}}$.}
		\ENDFOR
		\STATE{\textbf{Output:} The optimal solutions ${\bf{d}}$, $\bm{\beta}$, and ${{\bf{f}}^M}$ are obtained.}
	\end{algorithmic}
\end{algorithm}

\section{Numerical Results} \label{sec:NumericalResults} 
\begin{table}[!t] \scriptsize
	\centering
	\caption {Simulation Parameters} 
	\begin{tabular}{l|l}
		\hline \hline
		\textbf{Parameter} & \textbf{Value}          
		\\ \hline
		Number of FAs at the BS, $M$                           
		&
		$4$
		\\ \hline
		Number of users, $N$                                                       
		&
		$3$  
		\\ \hline
		Number of channel paths for each user, $L_n$ 
		&  
		$3$ 
		\\ \hline
		Carrier wavelength, $\lambda$
		&
		$0.1$ m
		\\ \hline
		Length of sides of receive area, $A$
		&
		$1.5\lambda$
		\\ \hline
		Minimum inter-FA distance, $d_0$
		&
		$\lambda$
		\\ \hline
		Data size of user $n$, ${D}_n$
		& 
		$\left[ {0.5 - 2} \right]$ KB
		\\ \hline
		Local CPU frequency of user $n$, ${f_n^{loc}}$                                                      
		&  
		$\left[ {0.8 - 1} \right]$ GHz	
		\\ \hline
		Elevation and azimuth AoAs, ${\theta _{n,l}}={\phi _{n,l}}$
		&
		$\left[ { - \frac{\pi }{2},\frac{\pi }{2}} \right]$
		\\ \hline
		Distance from users to BS, ${\cal X}_{n}$
		&
		$\left[ {20,100} \right]\;{\text{m}}$
		\\ \hline
		Channel gain at the reference distance, $\rho$
		&
		$-40$ dB
		\\ \hline
		Path loss exponent, $\alpha$
		&
		2.8
		\\ \hline
		Maximum CPU frequency of MEC server, ${{\bar f}_M}$                                                           
		& 
		$10$ GHz 
		\\ \hline
		Transmit power for each user, ${p_n}$                                                
		& 
		$30$ dBm 
		\\ \hline
		Noise power spectrum density, ${{\sigma ^2}}$                                                
		& 
		$-174$ dBm/Hz
		\\ \hline
	\end{tabular}
	\label{SimulationParameters}
\end{table}
\begin{figure*}[!t]
	\centering 
	\subfigure[Antenna Position VS Iterations]{ 
		\label{subfig:AntennaPosition} 
		\includegraphics[width=0.28\linewidth]{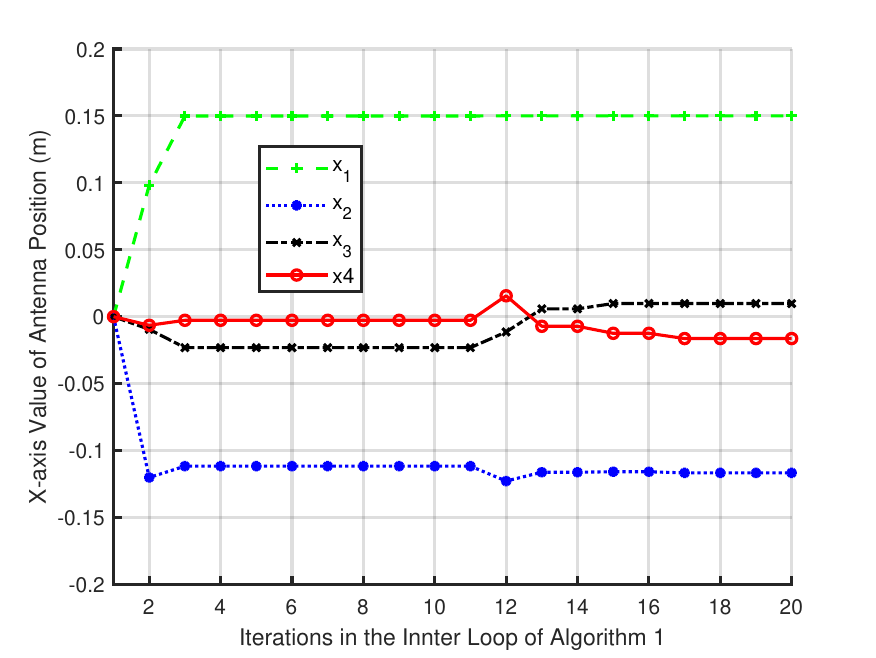}
	} 
	\subfigure[Total Latency VS Iterations]{ 
		\label{subfig:TotalLatency}\hspace{-5mm} 
		\includegraphics[width=0.28\linewidth]{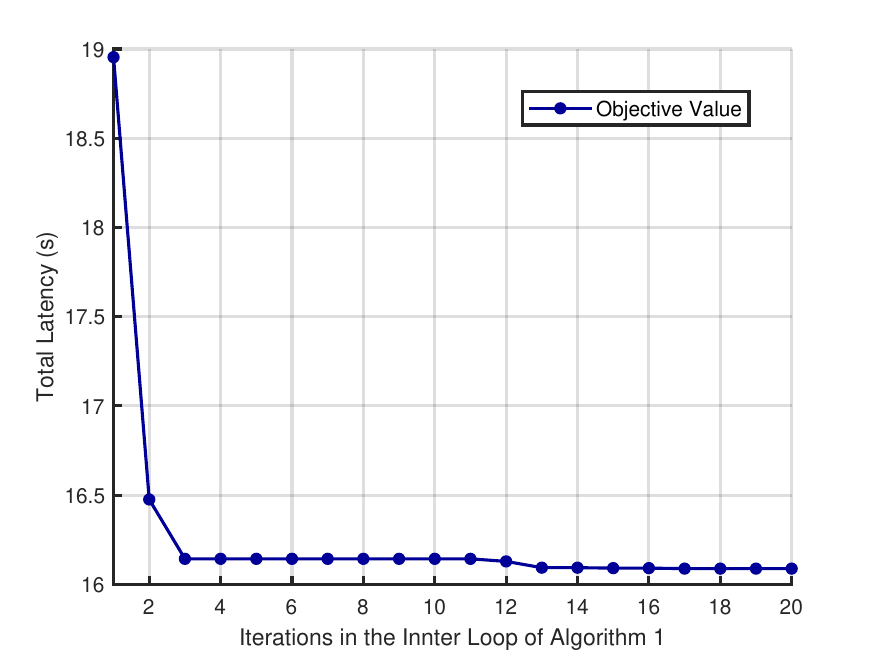}
	} 
	\subfigure[Offloading Ratio VS Iterations]{ 
		\label{subfig:OffloadingRatio}\hspace{-5mm} 
		\includegraphics[width=0.28\linewidth]{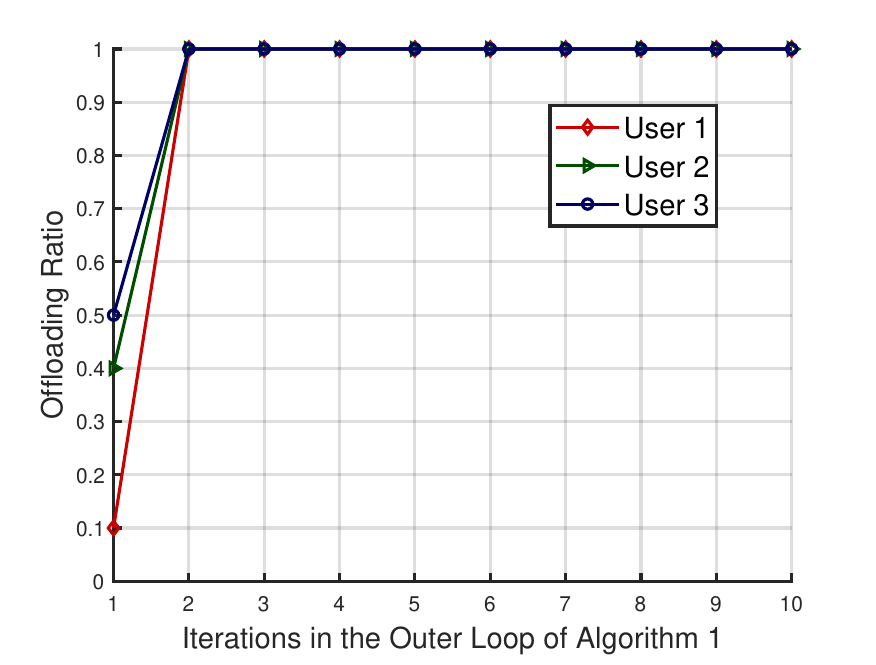}
	} 
	\caption{\small Convergence performance of the proposed alternating iterative algorithm.}
	\label{fig:Convergence} 
\end{figure*}
\begin{figure}
	\centering
	\includegraphics[width=0.65\linewidth]{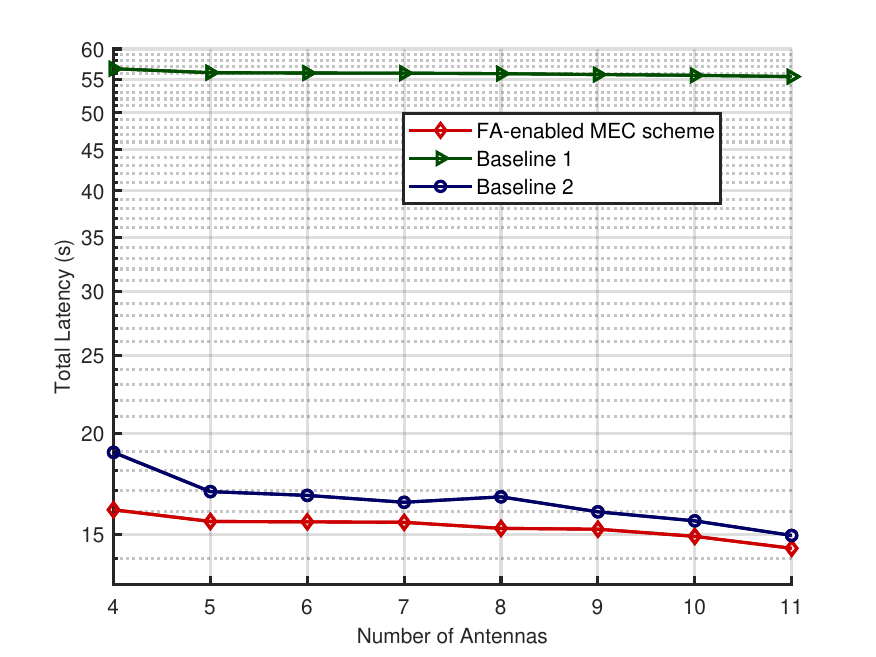}
	\caption{\small Comparison of our proposed FA-enabled MEC scheme with Baseline 1 and Baseline 2.}
	\label{fig:ComparisonTotalLatency}
\end{figure}
In this section, we first verify the convergence of the IPPOS-based alternating iterative algorithm. Then, we compare the total latency of the proposed alternating iterative algorithm with those of two baseline schemes. Each element in the path-response vector follows an independent and identically distributed circularly symmetric complex Gaussian distribution \cite{xiao2023multiuser}, denoted as ${g_{n,l}} \sim \mathcal{C}\mathcal{N}\left( {0,{{\rho \mathcal{X}_n^{ - \alpha }} \mathord{\left/{\vphantom {{\rho \mathcal{L}_n^{ - \alpha }} L}} \right. \kern-\nulldelimiterspace} L}} \right)$ for $n \in {\mathcal{N}}$ and $l \in {\mathcal{L}_n}$, where $\mathcal{X}_n$ is the distance from user $n$ to BS, $\rho$ denotes the channel gain at a standard reference distance of one meter, and $\alpha$ is the path loss exponent. The main simulation parameters are listed in Table~\ref{SimulationParameters}. Importantly, we should note that certain system parameters may vary across different simulation scenarios.

For ease of explanation, we consider a simple example involving 4 FAs at the BS, 3 users, and 3 channel paths for each user in the FA-enabled MEC network. Fig.~\ref{fig:Convergence} illustrates the convergence performance of the proposed IPPSO-based alternating iteration algorithm. Fig.~\ref{subfig:AntennaPosition} and Fig.~\ref{subfig:TotalLatency} detail the convergence process if the inner loop of Algorithm~\ref{Algorithm_IPPSO}, while Fig.~\ref{subfig:OffloadingRatio} depicts the convergence trend of the outer loop of Algorithm~\ref{Algorithm_IPPSO}. As shown in Fig.~\ref{subfig:AntennaPosition}, after exceeding 18 iterations of the PSO algorithm, the X coordinate values of the four antennas reach a stable state. Similarly, Fig.~\ref{subfig:TotalLatency} demonstrates that the overall system delay begins to steady after surpassing 18 iterations of the PSO process. Furthermore, Fig.~\ref{subfig:OffloadingRatio} indicates that the offloading ratio from users to the MEC server converges swiftly, requiring merely two iterations. In summary, Fig.~\ref{fig:Convergence} fully verifies that the proposed IPPSO-based alternating iteration algorithm has good convergence performance.

Fig.~\ref{fig:ComparisonTotalLatency} shows the comparison of the total latency between the proposed FA-enabled MEC scheme and two baseline schemes. In Baseline 1, users process all computations locally without offloading to the MEC server, and the BS's antennas are fixed at the coordinate origin. Baseline 2 allows users to offload computing tasks to the MEC server, yet still constrains the BS antennas to the origin point. The results show that Baseline 1 has the highest total latency, mainly due to the long delay caused by users relying only on local computing. Conversely, our proposed FA-enabled MEC scheme exhibits lower total latency than Baseline 2. Numerical results also reveal that as the number of BS's antennas increases, the total latency of all schemes shows a downward trend. Compared with Baseline 1 and Baseline 2 with fixed antenna positions, the FA-enabled MEC scheme effectively improves the channel quality and increases the transmission rate by jointly optimizing all FA positions in a continuous spatial field, thereby significantly reducing the total delay. This finding highlights the potential of FA technology to optimize the performance of MEC systems, especially its important role in reducing computing and communication delays.   

\section{Conclusion} \label{sec:Conclusion}
In this letter, we have introduced a novel FA-enabled MEC system model. In this FA-enabled MEC network, we formulated the optimization problem focusing on jointly optimizing the offloading ratio, CPU frequency, and antenna positioning to minimize the total latency of all users. To tackle the inherent challenges of this problem, we proposed an IPPSO-based alternating iteration algorithm to obtain optimal solutions. Numerical results indicate that the proposed IPPSO-based algorithm has a fast convergence rate. Compared with the two baselines, the proposed FA-enabled MEC scheme 
has more advantages in reducing the total delay of all users. In summary, the integration of FA technology into MEC systems utilizes the mobility of FAs within a local domain at the BS to enhance channel conditions and thus improve the efficiency of computational task offloading and processing. This demonstrates significant potential for the FA in the MEC system. 

\bibliographystyle{IEEEtran}
\bibliography{IEEEabrv,MAMEC}

\begin{thebibliography}{10}
\providecommand{\url}[1]{#1}
\csname url@samestyle\endcsname
\providecommand{\newblock}{\relax}
\providecommand{\bibinfo}[2]{#2}
\providecommand{\BIBentrySTDinterwordspacing}{\spaceskip=0pt\relax}
\providecommand{\BIBentryALTinterwordstretchfactor}{4}
\providecommand{\BIBentryALTinterwordspacing}{\spaceskip=\fontdimen2\font plus
\BIBentryALTinterwordstretchfactor\fontdimen3\font minus
  \fontdimen4\font\relax}
\providecommand{\BIBforeignlanguage}[2]{{%
\expandafter\ifx\csname l@#1\endcsname\relax
\typeout{** WARNING: IEEEtran.bst: No hyphenation pattern has been}%
\typeout{** loaded for the language `#1'. Using the pattern for}%
\typeout{** the default language instead.}%
\else
\language=\csname l@#1\endcsname
\fi
#2}}
\providecommand{\BIBdecl}{\relax}
\BIBdecl

\bibitem{mao2017survey}
Y.~Mao, C.~You, J.~Zhang, K.~Huang, and K.~B. Letaief, ``A survey on mobile
  edge computing: The communication perspective,'' \emph{IEEE Commun. Surv.
  Tut.}, vol.~19, no.~4, pp. 2322--2358, Aug. 2017.

\bibitem{zuo2023survey}
Y.~Zuo, J.~Guo, N.~Gao, Y.~Zhu, S.~Jin, and X.~Li, ``A survey of blockchain and
  artificial intelligence for {6G} wireless communications,'' \emph{IEEE
  Commun. Surv. Tut.}, vol.~25, no.~4, pp. 2494--2528, Sep. 2023.

\bibitem{abbas2017mobile}
N.~Abbas, Y.~Zhang, A.~Taherkordi, and T.~Skeie, ``Mobile edge computing: A
  survey,'' \emph{IEEE Internet Things J.}, vol.~5, no.~1, pp. 450--465, Sep.
  2017.

\bibitem{wong2023fluid27}
K.-K. Wong, W.~K. New, X.~Hao, K.-F. Tong, and C.-B. Chae, ``Fluid antenna
  system—part {I}: Preliminaries,'' \emph{IEEE Commun. Lett.}, vol.~27,
  no.~8, pp. 1919--1923, Aug. 2023.

\bibitem{zhu2023movable1}
L.~Zhu, W.~Ma, B.~Ning, and R.~Zhang, ``Movable-antenna enhanced multiuser
  communication via antenna position optimization,'' \emph{IEEE Trans. Wireless
  Commun.}, 2023, Early Access.

\bibitem{ramirez2024new}
P.~Ramirez-Espinosa, D.~Morales-Jimenez, and K.-K. Wong, ``A new spatial
  block-correlation model for fluid antenna systems,'' \emph{arXiv preprint
  arXiv:2401.04513}, 2024.

\bibitem{ye2023fluid}
Y.~Ye, L.~You, J.~Wang, H.~Xu, K.-K. Wong, and X.~Gao, ``Fluid antenna-assisted
  {MIMO} transmission exploiting statistical {CSI},'' \emph{IEEE Commun.
  Lett.}, vol.~28, no.~1, pp. 223--227, Jan. 2024.

\bibitem{zhu2024index}
J.~Zhu, G.~Chen, P.~Gao, P.~Xiao, Z.~Lin, and A.~Quddus, ``Index modulation for
  fluid antenna-assisted {MIMO} communications: System design and performance
  analysis,'' \emph{IEEE Trans. Wireless Commun.}, 2024, Early Access.

\bibitem{hu2024fluid}
G.~Hu, Q.~Wu, K.~Xu, J.~Ouyang, J.~Si, Y.~Cai, and N.~Al-Dhahir, ``Fluid
  antennas-enabled multiuser uplink: A low-complexity gradient descent for
  total transmit power minimization,'' \emph{IEEE Commun. Lett.}, vol.~28,
  no.~3, pp. 602--606, Mar. 2024.

\bibitem{ghadi2024performance}
F.~R. Ghadi, K.-K. Wong, W.~K. New, H.~Xu, R.~Murch, and Y.~Zhang, ``On
  performance of {RIS}-aided fluid antenna systems,'' \emph{arXiv preprint
  arXiv:2402.16116}, 2024.

\bibitem{wong2022extra}
K.-K. Wong, K.-F. Tong, Y.~Chen, and Y.~Zhang, ``Extra-large {MIMO} enabling
  slow fluid antenna massive access for millimeter-wave bands,''
  \emph{Electron. Lett.}, vol.~58, no.~25, pp. 1016--1018, Nov. 2022.

\bibitem{hu2023movable}
G.~Hu, Q.~Wu, K.~Xu, J.~Ouyang, J.~Si, Y.~Cai, and N.~Al-Dhahir,
  ``Movable-antenna array enabled multiuser uplink: A low-complexity gradient
  descent for total transmit power minimization,'' \emph{arXiv preprint
  arXiv:2312.05763}, 2023.

\bibitem{xiao2023multiuser}
Z.~Xiao, X.~Pi, L.~Zhu, X.-G. Xia, and R.~Zhang, ``Multiuser communications
  with movable-antenna base station: Joint antenna positioning, receive
  combining, and power control,'' \emph{arXiv preprint arXiv:2308.09512}, 2023.

\end{thebibliography}
\end{document}